\documentclass{ws-procs975x65}

\begin{document}

\title{Defrosting the Big Freeze quantum mechanically?}

\author{Mariam  Bouhmadi-L\'opez$^1$} 
\address{Centro Multidisciplinar de Astrof\'{\i}sica - CENTRA,
 Departamento de F\'{\i}sica,\\ Instituto Superior T\'ecnico,
Av. Rovisco Pais 1, 1096 Lisboa, Portugal}

\vspace*{-0.3cm}\author{Claus Kiefer$^2$ and Barbara Sandh\"ofer$^3$}
\address{\mbox{Institut f\"ur Theoretische Physik, Universit\"{a}t zu
K\"{o}ln, Z\"{u}lpicher Strasse 77, 50937 K\"{o}ln, Germany}}

\vspace*{-0.3cm}\author{Paulo Vargas Moniz$^4$}

\address{Departamento de F\'{\i}sica, Universidade da Beira
Interior, Rua Marqu\^{e}s d'Avila e Bolama,\\ 6200 Covilh\~{a},
Portugal\\\vspace*{0.5cm}E-mails: \mbox{$^1$mariam.bouhmadi@ist.utl.pt}, \mbox{$^2$kiefer@thp.uni-koeln.de}, \mbox{$^3$bs324@thp.uni-koeln.de}, \mbox{$^4$pmoniz@ubi.pt}}

\begin{abstract}
The big freeze is a singularity that shows up in some dark energy models. We address here its possible avoidance in quantum cosmology, more precisely in the framework of quantum geometrodynamics.
\end{abstract}


\bodymatter

\section{Motivation}

Over the last years it has been realised that several dark energy related singularities can show up in the future evolution of the universe \cite{Nojiri}. Among them is the big freeze singularity \cite{BouhmadiLopez:2006fu}, a subclass of type III singularity \cite{Nojiri}, which takes place at finite scale factor and finite cosmic time in a (flat) Friedmann-Lema\^{\i}tre-Robertson-Walker (FLRW) universe. 

Our main objective here lies not so much in the observational significance of some of these dark energy models but in their relevance for understanding the quantisation of gravity. In fact, the search for a consistent theory of quantum gravity is among the main open problems in theoretical physics \cite{OUP}.

The future big freeze singularity can appear in a FLRW universe filled with a (phantom) generalised Chaplygin gas (GCG). To be able to study the quantum behaviour, the GCG has to be mimicked by a fundamental field because also the matter part should have its own degrees of freedom. We shall carry out our quantisation in the geometrodynamical framework; i.e. using the Wheeler-DeWitt equation. 

\section{The big-freeze singularity induced by a scalar field}

The GCG can be mimicked by a minimally coupled scalar field, $\phi$. Here, as we are dealing with a phantom GCG, whose equation of state is $P=-A/\rho^{\beta}$ where $A>0$ and $1+\beta<0$, the energy density and pressure of $\phi$ read
\begin{equation}
  \rho_\phi= -\frac12 \dot\phi^2 + V(\phi)\ , \quad
  p_\phi=-\frac12 \dot\phi^2 - V(\phi),
\end{equation}
while the potential can be expressed as 
\begin{equation}
  V(\phi)=V_{0}\left[\frac{1}{\sin^{\frac{2\beta}{1+\beta}}\left(\frac{\sqrt{3}}{2}\kappa|1+\beta||\phi|\right)}+
    \sin^{\frac{2}{1+\beta}}\left(\frac{\sqrt{3}}{2}\kappa|1+\beta||\phi|\right)
  \right].
  \label{vphi2}
\end{equation}
In the previous equation $V_{0}=A^{\frac{1}{1+\beta}}/2$, $\kappa^2=8\pi G$ and
$0<(\sqrt{3}/{2})\kappa|1+\beta||\phi|\leq \pi/2$. The big freeze singularity takes place when $\phi\rightarrow 0$ which corresponds to the scale factor reaching its maximum allowed classical value.

\section{The Wheeler-DeWitt equation}

Next, we shall describe the quantisation of the classical scenario described above. This will be carried out using the Wheeler-DeWitt equation which in our case reads \cite{paper}
\begin{equation}
  \frac{\hbar^2}{2}\left(\frac{\kappa^2}{6}\frac{\partial^2}{\partial\alpha^2}+\frac{\partial^2}{\partial\phi^2}\right)
  \Psi\left(\alpha,\phi\right)+
  a_{\rm{max}}^6e^{6\alpha}V(\phi)\Psi\left(\alpha , \phi\right)\nonumber =0,
  \label{WdW1}
\end{equation}
where $V(\phi)$ is given  in (\ref{vphi2}). We have introduced the  variable
$\alpha:=\ln\left(\frac{a}{a_{\rm{max}}}\right)$ where $a_{\rm{max}}$
corresponds to the location of the singularity. In the following, we shall use $\tilde{a}:=\frac{a}{a_0}$
instead of $a$. For simplicity, we shall drop the tilde. To solve this
equation, we make a Born--Oppenheimer-type of \emph{ansatz}
\begin{equation} \Psi(\alpha,\phi)=\varphi_k(\alpha,\phi)C_k(\alpha)\ . \end{equation}
Furthermore, we require $\varphi_k$ to satisfy (close to the singularity)
\begin{equation} \label{varphieqn} \varphi_k^{\prime \prime}+\left[
- k^2+\tilde{V}_{\alpha}\vert\phi\vert^{-\frac{2\beta}{1+\beta}}\right]\varphi_k=0\
, \end{equation}
where $^\prime$ denotes a derivative with respect to $\phi$ and 
\begin{equation}
\tilde{V}_{\alpha}:=\frac{2}{\hbar^2}a_{\rm{max}}^6e^{6\alpha}V_{-1}\left[\frac{\sqrt{3}\kappa}{2}\vert1+\beta\vert\right]^{-\frac{2\beta}{1+\beta}}.
\end{equation}
This equation is formally the same as the radial part of the stationary Schr\"odinger equation for an {\em attractive}
potential of inverse power. For simplicity, we will restrict to the
inverse square potential which is realized for $\vert\beta\vert\gg 1$, where $\beta$ is chosen such that $\vert1+\beta\vert\vert\phi\vert$ is still small.

Concerning the gravitational part of the wave function, it satisfies 
\begin{equation}\label{Ckeqn}
\left(\frac{\kappa^2}{6}\ddot
    C_k+ k^2C_k\right)\varphi_k=0\, , 
\end{equation}
where a dot denotes derivative with respect to $\alpha$.  To deduce the previous equation, we have assumed that 
the change in  the matter part does not influence the gravitational part; the matter part simply contributes its energy through $k^2$.

\section{Singularity avoidance}

Following the approach we have presented in the previous section, the wave function can be deduced analytically (for $\vert\beta\vert\gg 1$). The matter part reads
\begin{equation}\label{generalvarphik}
\varphi_k(\alpha,\vert\phi\vert)=\sqrt{\vert\phi\vert}\left[c_1\mathrm{J}_\nu(i k\vert\phi\vert)+c_2\mathrm{Y}_\nu(i k\vert\phi\vert)\right]\
, \end{equation} 
where $\nu:=\sqrt{\frac14-\tilde{V}_\alpha}$. On the other hand, the gravitational part  of the wave function reads
\begin{equation}\label{gravsolution}
C_k(\alpha)=b_1e^{\mathrm{i}\frac{\sqrt6k}{\kappa}\alpha}+b_2e^{-\mathrm{i}\frac{\sqrt6k}{\kappa}\alpha}\
.  \end{equation}

It can be shown that the wave function vanishes close to the big freeze singularity ($\phi\rightarrow 0$) (see \cite{paper} for details). Therefore, we can conclude that the big freeze singularity is avoided. Notice that we have not imposed any boundary condition on the wave function of the universe. 

\section{Conclusions and further comments}

We have shown that the big freeze classical singularity (for the model we have presented) can be avoided in quantum cosmology because the wave function always vanishes at the singularity.

As a last comment, we would like to highlight that if one wants
to construct wave packets that follow classical trajectories with
turning point in configuration space \cite{OUP}, one has to require that the wave packet decays in the classically forbidden region. This is just the standard
quantum mechanical treatment of classically forbidden regions. In
general, out of solutions to the Wheeler--DeWitt equation which grow
in the classically forbidden region, no wave packet can be
constructed that follows the classical path. In order to make
connection with the underlying classical theory, we can therefore
use as condition that the wave function decreases in the
classically forbidden region. This type of boundary condition can be imposed in the model we have discussed. A more detailed discussion of these, and other, models can be found in the corresponding publication \cite{paper}.

\section*{Acknowledgements}
\vspace*{-0.2cm}
\mbox{M.B.L. is supported by FCT (Portugal) through the fellowship FCT/BPD/26542/2006}. This research work was supported by  the grant FEDER-POCI/P/FIS/57547/2004.

\end{document}